\documentclass[journal]{IEEEtran}

\ifCLASSINFOpdf
  \usepackage[pdftex]{graphicx}
\else
\fi

\usepackage{cite}
\usepackage{amsmath,amssymb}
\usepackage{graphicx}
\usepackage{hyperref}
\usepackage{multirow}
\usepackage{algorithm}
\usepackage{algorithmic}
\usepackage{booktabs}
\usepackage{tikz}
\usepackage{array}
\usepackage{hyperref}
\usepackage{pifont}
\newcommand{\cmark}{\ding{51}} 
\newcommand{\xmark}{\ding{55}} 
\usepackage{caption}

\definecolor{lightgreen}{RGB}{210, 255, 210}
\definecolor{lightred}{RGB}{255, 210, 210}
\definecolor{lightblue}{RGB}{210, 230, 255}
\definecolor{pinktext}{RGB}{180, 0, 90}

\begin{document}

\title{Efficient Routing of Inference Requests across \\ LLM Instances in Cloud-Edge Computing}

\author{Shibo Yu$^{1}$, Mohammad Goudarzi$^{1}$, and Adel Nadjaran Toosi$^{2}$ 
\\
$^{1}$The Faculty of Information Technology, Monash University, Australia \\
$^{2}$School of Computing and Information Systems, The University of Melbourne, Australia%

}

\markboth{}%
{Yu \MakeLowercase{et al.}: Efficient Routing of Inference Requests across \\ LLM Instances in Cloud-Edge Computing}

\maketitle


\begin{abstract}
The rising demand for Large Language Model (LLM) inference services has intensified pressure on computational resources, resulting in latency and cost challenges. This paper introduces a novel routing algorithm based on the Non-dominated Sorting Genetic Algorithm II (NSGA-II) to distribute inference requests across heterogeneous LLM instances in a cloud-edge computing environment. Formulated as a multi-objective optimization problem, the algorithm balances response quality, response time, and inference cost, adapting to request heterogeneity (e.g., varying complexity and prompt lengths) and node diversity (e.g., edge vs. cloud resources). This adaptive routing algorithm optimizes performance under dynamic workloads. We benchmark the approach using a testbed with datasets including Stanford Question Answering Dataset (SQuAD), Mostly Basic Python Problems (MBPP), Hella Situations With Adversarial Generations (HellaSwag), and Grade School Math 8K (GSM8K). Experimental results show our solution, compared to the baselines, preserves 95.2\% of Cloud-Only response quality with slight
latency increase, while reducing inference cost by 34.9\%. These findings validate the algorithm's effectiveness for scalable LLM deployments.
\end{abstract}

\begin{IEEEkeywords}
large language model, inference request routing, meta-heuristic, cloud-edge computing
\end{IEEEkeywords}

\section{Introduction}

In recent years, Large Language Models (LLMs) have demonstrated remarkable capabilities in understanding and generating human language, enabling their widespread adoption in domains such as virtual assistants, programming support, and content generation \cite{fan2024bibliometric, Wang2023b, Adiwardana2020, Li2023}. However, as the demand for LLM-based services surges, so does the volume of inference requests, placing tremendous pressure on underlying computational infrastructure, particularly Graphics Processing Units (GPUs). This increased workload often leads to bottlenecks in processing, prolonged response times, and degraded quality of service \cite{agrawal2024taming}.

To address the performance and scalability issues associated with high-volume LLM inference workloads, researchers have explored various strategies, most notably optimizing request scheduling at the level of individual LLM instances \cite{Yu2022, zhong2024distserve, Patel2024}. These techniques often integrate request batching and prioritization mechanisms to improve throughput and minimize queuing delays. Despite their effectiveness in isolated settings, such solutions remain confined to single-instance scenarios and fall short when extended to multi-instance deployments. As demand continues to grow, single-instance optimization becomes insufficient, making it imperative to develop methods capable of efficiently distributing requests across multiple LLM instances \cite{xu2024deploying}.

Recent studies have highlighted the importance of cross-instance routing and scheduling to enhance the scalability and responsiveness of LLM services \cite{Jain2024}. However, this task presents multiple challenges. First, LLM inference requests are inherently heterogeneous, varying significantly in size, complexity, and latency sensitivity. Assigning diverse workloads to a single instance without intelligent scheduling can lead to resource contention, increased inference time, and poor user experience. Second, the scheduling problem becomes even more complex when LLMs are deployed across a hybrid infrastructure that includes both cloud and edge environments \cite{chen2024transforming}.

The adoption of cloud-edge collaborative architectures introduces new opportunities as well as challenges for LLM inference optimization \cite{Friha2024, Goudarzi2023DDRL}. On one hand, edge devices offer the advantage of proximity to users, which helps reduce end-to-end latency. On the other hand, they are typically resource-constrained and may not be suitable for handling complex inference requests. Cloud servers, in contrast, provide high computational power but suffer from increased latency due to network overhead. Efficiently distributing inference workloads across such heterogeneous environments requires a nuanced understanding of task characteristics, hardware capabilities, and latency trade-offs \cite{zheng2025review}. A naive load-balancing strategy is unlikely to yield satisfactory results in this context.

Furthermore, existing research on inference request scheduling has largely overlooked two critical aspects: (1) the potential of leveraging cloud-edge collaboration for LLM request inference \cite{ding2024hybrid, Feng2024, Stripelis2024, Zhao2024}, and (2) the use of advanced decision-making algorithms such as meta-heuristic to route inference requests. Meta-heuristic like Non-dominated Sorting Genetic Algorithm II (NSGA-II) offers a model-free learning paradigm capable of continuously improving routing policies through trial-and-error interaction with the environment \cite{ma2023comprehensive}. This makes it a promising candidate for dynamic scheduling in real-world, data-scarce scenarios.

In this work, we propose a novel LLM inference task routing algorithm. The algorithm adaptively routes user queries across edge and cloud LLM instances using a router trained via the NSGA-II. The key idea is to exploit the strengths of both edge and cloud environments—using small, resource-efficient LLMs on edge devices for lightweight requests and powerful cloud-based LLMs for more complex queries—thereby achieving a balance among responsiveness, cost-effectiveness and response quality.

Our key contributions are summarized as follows:

\begin{itemize}
    \item We formulated the inference task routing problem as a multi-objective optimization problem in terms of inference quality, inference cost and response time.
    \item We design a novel routing algorithm based on NSGA-II to dynamically route inference requests across heterogeneous LLM instances in a cloud-edge environment.
    \item We develop a set of performance testing tools for distributed computing environment, containing datasets commonly used for LLM performance testing, such as Stanford Question Answering Dataset (SQuAD) \cite{rajpurkar2016squad}, Mostly Basic Python Problems (MBPP) \cite{austin2021program}, Hella Situations With Adversarial Generations  (HellaSwag) \cite{zellers2019hellaswag}, Grade School Math 8K (GSM8K) \cite{zhang2024careful}. Previously, similar performance testing tools were designed for testing on a single node or in the cloud \cite{hu2024routerbench, wang2024bench}, while these performance testing tools have not been adapted for distributed computing environments.
    \item In a real deployment environment, our experimental results validate the effectiveness of our approach in reducing average latency and inference cost while preserving inference accuracy.
\end{itemize}

The remainder of this paper is organized as follows. Section 2 reviews related work on cloud and edge environment, LLMs and transformer architecture, meta-heuristic and LLM Request Routing. Section 3 describes the problem formulation. Section 4 describes the design of our algorithm. Section 5 presents the implementation, datasets, experimental setup, and the result. Section 6 discusses the results and analysis. Finally, Section 7 concludes the paper and outlines directions for future work.

\section{Background}

In this section, we provide the required background information for this research, which is cloud and edge collaboration, LLMs and transformer architecture, and metaheuristic in multi-objective optimization. Also, we conduct a comprehensive study of the related work and highlights our contributions.
 
\subsection{Cloud and Edge Collaboration}


The cloud–edge architecture combines centralized cloud infrastructure with decentralized edge nodes to optimize data processing and service delivery \cite{Duan2020, Goudarzi2021FogBus2Arxiv, Wang2024ReinFog}. Cloud servers provide vast computational resources, scalability, and centralized data management, making them ideal for intensive tasks such as big data analytics and model training. However, their geographic distance from end users often introduces latency, which can negatively impact real-time applications \cite{shukla2023improving}. Conversely, edge nodes—deployed closer to data sources—enable rapid, low-latency responses by processing data locally and reducing network congestion and transmission delays \cite{Mach2017, Cao2020}. This hybrid paradigm supports context-aware workload placement by dynamically allocating tasks based on resource availability, latency requirements, and network conditions. For example, in the industrial Internet of Things, edge nodes preprocess sensor data for real-time monitoring, while the cloud aggregates insights for long-term analytics \cite{wang2023distributed, Wang2025TFDDRL}. Similarly, in smart cities, edge nodes manage local traffic data, and the cloud optimizes city-wide infrastructure \cite{trigka2025edge}. In LLM services, adaptive task scheduling in cloud-edge environments entails mapping inference requests to nodes hosting appropriate models, thereby balancing inference quality, inference cost, and response time. This distributed approach opens up opportunities for efficient, cost-effective LLM inference and enhances user experience across heterogeneous computing environments.

\subsection{LLMs and Transformer Architecture}
Large Language Models (LLMs) have become pivotal in modern artificial intelligence, demonstrating state-of-the-art performance across a spectrum of NLP tasks. Predominantly based on the Transformer architecture, LLMs utilize self-attention mechanisms to model token dependencies across arbitrary sequence distances \cite{Vaswani2017}. This innovation significantly surpasses RNN and LSTM models in handling long-range context \cite{Tang2018, Zhao2020}. A Transformer typically consists of multiple layers of self-attention and feedforward networks, enabling parallelized computation during the prefill phase and autoregressive token generation during the decode phase \cite{Lin2024}. However, LLM inference is computationally expensive, with memory and latency bottlenecks, particularly in the decode stage. To alleviate such challenges, works like Orca \cite{Yu2022} and vAttention \cite{prabhu2025vattention} propose iteration-aware scheduling and memory-optimized attention mechanisms, respectively, enabling more efficient resource management across iterations and long sequences.

\subsection{Metaheuristic in Multi-objective Optimization}
Metaheuristics are high‑level, problem‑independent strategies for finding near‑optimal solutions to complex optimization problems where exact methods are infeasible. They iteratively improve candidate solutions by emulating mechanisms drawn from natural, physical, or social processes. Formally, a metaheuristic $MH$ can be represented as
\[
MH = (O, A, R_c, R_i, R_o),
\]
where $O$ is a set of metaheuristic methodologies, $A$ a set of generic algorithms, $R_c$ the internal relations, and $R_i$ and $R_o$ the input and output relations, respectively.

A central design challenge is balancing exploration and exploitation: exploration spreads search to unvisited regions, while exploitation refines promising areas. Techniques such as adaptive parameter control, hybridization with local search, and adaptive memory programming have been proposed to manage this trade-off effectively \cite{crepinscek2013exploration}, \cite{mahdavi2015metaheuristics}.

Metaheuristics have also been extensively extended to multi-objective optimization problems \cite{hussain2019metaheuristic}. These multi-objective metaheuristics introduce Pareto dominance, external archives, and diversity maintenance to approximate a well-distributed Pareto front \cite{taha2020methods}. Notable multi-objective algorithms include NSGA-II, a fast elitist genetic algorithm that employs non-dominated sorting and crowding distance to maintain a diverse Pareto front \cite{Deng2021FogBus2}; Multi-Objective Particle Swarm Optimization, which uses Pareto dominance to update personal and global bests and relies on an external archive to guide the swarm toward diverse optimal solutions \cite{lalwani2013comprehensive}; and Multi-Objective Ant Colony Optimization, integrating Pareto-based pheromone updates and multiple pheromone matrices to balance convergence and diversity \cite{awadallah2025multi}. Their proficiency in optimizing multiple objectives positions them as promising solutions for inference request routing in cloud-edge systems.

\subsection{Related Work}

Recent advancements in LLM inference focus on optimizing throughput, latency, and resource efficiency for large-scale deployments.

Orca \cite{Yu2022} employs iteration-level scheduling and selective batching for Transformer-based models, enabling low-latency, high-throughput inference. It schedules model iterations, checking for completed requests after each, improving responsiveness over request-level scheduling. Sarathi-Serve \cite{agrawal2024taming} balances throughput and latency using chunked-prefills and stall-free batching, processing prompt and decode phases concurrently. Tested on models like Mistral-7B, it reduces pipeline stalls and outperforms Orca in multi-GPU settings. FastServe \cite{Wu2023} uses preemptive scheduling and MLFQ to minimize job completion time, with token-level preemption and proactive memory management. It outperforms Orca under high loads.

DistServe \cite{zhong2024distserve} disaggregates prefill and decoding phases, scheduling them on separate resources to optimize goodput. This decoupling improves resource utilization and throughput while maintaining competitive latency. Splitwise \cite{Patel2024} builds on phase splitting, similar to DistServe, by separating immediate computation and token generation across different machines. Using optimized data transfers over Infiniband, it achieves low latency and high throughput. vAttention \cite{prabhu2025vattention} dynamically manages memory to reduce overhead in LLM inference, using OS-level demand paging instead of vLLM’s PagedAttention \cite{kwon2023efficient}. This approach enhances GPU resource efficiency and throughput. The SSJF algorithm \cite{qiu2024efficient} predicts output lengths with a BERT-based proxy model, prioritizing shorter jobs to reduce blocking. It surpasses FCFS scheduling in job completion time and throughput.

While inference optimization improves throughput and latency, with the growing diversity of LLMs—varying in size, cost, latency, and domain specialization—the need for intelligent routing of inference requests has emerged as a key research area. Rather than adopting a one-size-fits-all model, recent systems propose routing frameworks that dynamically assign queries to the most suitable LLM to optimize metrics such as quality, latency, or cost \cite{ong2024routellm}.

It improves resource utilization over static methods. Early methods such as Tabi \cite{Wang2023a} routes simple queries to smaller models and complex ones to larger LLMs, using attention-based pruning to minimize latency and accuracy loss. Jha et al. \cite{Jha2024} use a DRL-based router to select models dynamically based on load, balancing quality and latency. Hybrid LLM \cite{ding2024hybrid} addressed this tradeoff by routing "easy" queries to small models and more complex ones to larger models, based on a predicted difficulty score. Although effective in binary (small-vs-large) settings, its applicability to multi-model systems is limited. To generalize beyond two models, Eagle \cite{Zhao2024} introduces a training-free router that combines global and local ELO-based ranking mechanisms. It infers model performance from historical feedback and achieves both scalability and real-time adaptability in high-throughput environments by avoiding costly retraining.

To address the routing trilemma—balancing inference cost, response quality, and throughput—PolyRouter \cite{Stripelis2024} treats LLMs as domain-specific experts and builds predictive classifiers that learn to route based on query semantics. Using soft labels derived from BERT similarity and negative log-likelihood, PolyRouter trains lightweight classifiers to predict the best LLM for each input. This design significantly improves efficiency and reduces cost compared to naive round-robin or static policies. In parallel, GraphRouter \cite{Feng2024} rethinks routing from a structural perspective, modeling tasks, queries, and LLMs as nodes in a heterogeneous graph. The system performs inductive edge prediction to estimate the cost and effectiveness of LLM-query pairs. By leveraging graph neural networks (GNNs), GraphRouter generalizes to unseen queries and new LLMs without retraining. This architecture captures nuanced relationships among tasks and models, offering a robust solution for evolving LLM ecosystems.

On the multi-model routing in cloud environments side, Intelligent Router \cite{Jain2024} targets low-latency serving in cloud clusters by using workload-aware reinforcement learning. It models request characteristics—such as prompt/decode length—and their impact on LLM instance-level batching. A lightweight predictor forecasts decoding time, while a RL-based policy routes requests to instances in a way that avoids mixing heavy and light tasks, achieving notable reductions in end-to-end latency.

Collectively, these works underscore a shift from static, monolithic serving architectures to adaptive, context-aware routing frameworks. As LLM-based services scale, efficient query routing becomes essential not only for cost savings but also for maximizing user experience and system sustainability. Despite recent advances, most approaches still focus on inference request routing in a single node or server cluster, which limits the great potential of cloud-edge collaboration. Integrating NSGA-II into LLM routers within cloud-edge contexts presents a promising direction, enabling adaptation to multi-objective trade-offs among response time, inference cost, and inference quality across cloud-edge architectures.

\begin{table*}[t]
\caption{Comparison of LLM Serving Optimization Approaches and Evaluation}
\label{tab:llm_serving_comparison}
\centering
\renewcommand{\arraystretch}{1.2}
\begin{tabular}{
    >{\raggedright\arraybackslash}p{1.65cm}  
    >{\centering\arraybackslash}p{1.1cm}  
    >{\centering\arraybackslash}p{1.1cm}  
    >{\centering\arraybackslash}p{1.15cm}  
    >{\centering\arraybackslash}p{1.5cm}  
    >{\centering\arraybackslash}p{0.8cm}  
    >{\centering\arraybackslash}p{0.7cm}  
    >{\centering\arraybackslash}p{0.7cm}  
    >{\raggedright\arraybackslash}p{1.6cm}  
    >{\raggedright\arraybackslash}p{1.6cm}  
    >{\raggedright\arraybackslash}p{1.5cm}  
}
\toprule
\textbf{Reference} &
\multicolumn{4}{c}{\textbf{Environmental Architecture}} &
\multicolumn{4}{c}{\textbf{Level of Optimization for LLM Serving}} &
\multicolumn{2}{c}{\textbf{Experimental Evaluation}} \\
\cmidrule(lr){2-5} \cmidrule(lr){6-9} \cmidrule(lr){10-11}
 & \textbf{Cloud/Edge} & \textbf{Multi Servers} & \textbf{\# LLMs Deployed} & \textbf{Heterogeneity LLMs} & \textbf{Inter Server} & \textbf{Inter LLM} & \textbf{Intra LLM} & \textbf{Technique} & \textbf{Experimental Environment} & \textbf{Evaluation Metrics} \\
\midrule
Orca \cite{Yu2022}                   & Cloud      & \cmark & Single           & NA & \xmark & \xmark & \cmark & Heuristic & Actual Deployment & Latency \& Throughput \\
Splitwise \cite{Patel2024}          & Cloud      & \cmark & Single           & NA & \xmark & \xmark & \cmark & Heuristic & Actual Deployment & Latency \& Throughput \\
Sarathi-Serve \cite{agrawal2024taming}           & Cloud      & \cmark & Single           & NA & \xmark & \xmark & \cmark & Heuristic & Actual Deployment & Latency \\
Distserve \cite{zhong2024distserve}          & Cloud      & \cmark & Single           & NA & \xmark & \xmark & \cmark & Heuristic & Actual Deployment & SLO \\
vAttention \cite{prabhu2025vattention}        & Cloud      & \xmark & Single           & NA & \xmark & \xmark & \cmark & VM Mechanism & Actual Deployment & Throughput \\
Qiu et al. \cite{qiu2024efficient}           & Cloud      & \xmark & Single           & NA & \xmark & \xmark & \cmark & Heuristic & Actual Deployment & JCT \& Throughput \\
FastServe \cite{Wu2023}             & Cloud      & \cmark & Single           & NA & \xmark & \xmark & \cmark & Heuristic & Actual Deployment & JCT \\
Jha et al. \cite{Jha2024}           & Cloud      & \xmark & Multiple         & \cmark & \xmark & \cmark & \xmark & DQN & Actual Deployment & Throughput \\
Tabi \cite{Wang2023a}               & Cloud      & \xmark & Multiple         & \cmark & \xmark & \cmark & \xmark & Heuristic & Actual Deployment & Latency \& Accuracy \\
Intelligent router \cite{Jain2024}  & Cloud      & NM     & Multiple         & \xmark & \xmark & \cmark & \xmark & DDQN & Actual Deployment & Latency \\
Polyrouter \cite{Stripelis2024}     & Cloud      & \xmark & Multiple         & \cmark & \xmark & \cmark & \xmark & Pre-trained BERT model & Actual Deployment & Response Quality \& Throughput \& Inference Cost \\
Hybrid LLM \cite{ding2024hybrid}    & Cloud      & \xmark & Multiple         & \cmark & \xmark & \cmark & \xmark & Pre-trained BERT model & Actual Deployment & Response Quality \& Inference Cost \\
Eagle \cite{Zhao2024}               & Cloud      & \xmark & Multiple         & \cmark & \xmark & \cmark & \xmark & Elo rating system & Actual Deployment & Response Quality \\
Graphrouter \cite{Feng2024}         & Cloud      & \xmark & Multiple         & \cmark & \xmark & \cmark & \xmark & Heterogeneous GNN & Actual Deployment & Response Quality \& Inference Cost \\
Routellm \cite{ong2024routellm}             & Cloud      & NM     & Use online LLMs  & \cmark & \xmark & \cmark & \xmark & Matrix Factorization, BERT Classifier & Actual Deployment & Response Quality \& Inference Cost \\
\textbf{Proposed Solution}              & Cloud-Edge & \cmark & Multiple         & \cmark & \cmark & \cmark & \xmark & NSGA-II & Actual Deployment & Response Time \& Inference Cost \& Inference Quality \\
\bottomrule
\multicolumn{11}{p{0.96\textwidth}}{\footnotesize NM: Not Mentioned, NA: Not Applicable, VM Mechanism: Virtual Memory Mechanism, DQN: Deep Q-Network, DDQN: Double DQN, DRL: Deep Reinforcement Learning, JCT: Job Completion Time, SLO: Service-Level Objective.}
\end{tabular}
\end{table*}



Table I is a comparative analysis of related work. The Reference column lists citation sources, while Cloud/Edge describes the system architecture. Number of Server indicate the server count. Heterogeneity LLMs specifies whether different types of LLMs are used, and Among Server and Among LLMs describe where scheduling or resource management technique is applied among servers or among LLMs. Technique details the specific technique used, Experimental Environment describes the setting, and Evaluation Metrics outlines performance criteria like latency and throughput. In our solution, multiple nodes are deployed in a cloud edge environment, with the cloud and edge nodes hosting different models. Our routing algorithm determines how to route each request to a specific model on a specific node.

\section{Problem Formulation}

We aim to optimize the routing of Large Language Model (LLM) inference requests across a cloud-edge architecture, balancing three objectives: response quality, inference cost, and response time. The problem is formulated as a multi-objective optimization task, where a routing policy assigns each inference request to a specific node and LLM instance to minimize latency and cost while maximizing response quality.

Let \( R = \{r_1, r_2, \dots, r_I\} \) denote the set of \( I \) inference requests, where each request \( r_i \) has attributes such as prompt length, complexity, and category. Let \( N = \{n_1, n_2, \dots, n_N\} \) denote the set of \( N \) nodes in the cloud-edge system, categorized into cloud nodes and edge nodes such that \( N = N_{\text{cloud}} \cup N_{\text{edge}} \) and \( N_{\text{cloud}} \cap N_{\text{edge}} = \emptyset \). Let \( M = \{m_1, m_2, \dots, m_K\} \) represent the global set of \( K \) Large Language Models (LLMs), where each model \( m_k \in M \) is characterized by its computational capacity, cost per token, and inference quality.

In a theoretical setting, we assume every node \( n_j \in N \) can deploy all models in \( M \). The solution space \( S \) for routing decisions is defined as the Cartesian product of nodes and models:
\[
S = N \times M = \{ (n_j, m_k) \mid n_j \in N, m_k \in M \},
\]
where each pair \( (n_j, m_k) \) represents the assignment of model \( m_k \) on node \( n_j \) to process a request. The size of this solution space is \( |S| = |N| \times |M| \), with \( |N| \) and \( |M| \) being the cardinalities of the node and model sets, respectively.

However, in practical deployments, hardware and resource constraints limit model availability across nodes. To capture this heterogeneity, we define a model deployment function \( \mathcal{M}: N \to 2^M \), which maps each node \( n_j \in N \) to a subset of models \( \mathcal{M}(n_j) \subseteq M \) that are actually deployed on it. Consequently, the solution space is refined as:
\[
S = \{ (n_j, m_k) \mid n_j \in N, m_k \in \mathcal{M}(n_j) \}.
\]

The routing policy \( \pi: R \to S \) maps each inference request \( r_i \in R \) to a node-LLM pair \( (n_j, m_k) \in S \), where \( m_k \in \mathcal{M}(n_j) \), ensuring that requests are assigned only to valid node-LLM combinations available in the system. The multi-objective optimization problem is defined as:
\begin{equation}
\min \left( \omega_1 RQ + \omega_2 C + \omega_3 RT \right)
\label{eq:objective}
\end{equation}
where \( RQ \) is the response quality metric, \( C \) is the inference cost metric, \( RT \) is the response time metric, and \( \omega_1, \omega_2, \omega_3 \in [0, 1] \) are weight coefficients satisfying \( \omega_1 + \omega_2 + \omega_3 = 1 \), used to balance the trade-offs among the objectives. Since NSGA-II is employed to find a Pareto-optimal set of solutions, these weights are used to aggregate objectives for evaluation but can be adjusted to explore the Pareto front.

The optimization is subject to the following constraints:
\begin{itemize}
    \item \textbf{Resource Constraint}: Each node \( n_j \) has limited computational resources (e.g., GPU memory, processing capacity), denoted by \( C_j \). The total resource demand of requests assigned to \( n_j \) must not exceed \( C_j \).
    \item \textbf{Assignment Constraint}: Each request \( r_i \) is assigned to exactly one node-LLM pair \( (n_j, m_k) \).
\end{itemize}

\subsection{\texorpdfstring{Response Quality Metric (\( RQ \))}{Response Quality Metric (RQ)}}

The response quality metric \( RQ \) quantifies the average quality of responses across all requests, defined as:
\begin{equation}
RQ = \frac{1}{I} \sum_{i=1}^I (1 - q(r_i))
\label{eq:rq}
\end{equation}
where \( q(r_i) \in [0, 1] \) is the quality score of the response for request \( r_i \), computed based on dataset-specific evaluation metrics. For example, responses for the MBPP dataset are evaluated using CodeBLEU \cite{ren2020codebleu}, while SQuAD responses use F1-score \cite{rajpurkar2016squad}, HellaSwag uses accuracy \cite{zellers2019hellaswag}, and GSM8K uses exact match \cite{zhang2024careful}. Since higher \( q(r_i) \) indicates better quality, minimizing \( RQ \) (i.e., maximizing \( q(r_i) \)) ensures high-quality responses across all requests.

\subsection{\texorpdfstring{Inference Cost Metric (\( C \))}{Inference Cost Metric (C)}}
The inference cost metric \( C \) represents the average cost per request, calculated as:
\begin{equation}
C = \frac{1}{I} \sum_{i=1}^I \left( \frac{l_i}{10^6} \cdot p_{m_k} \right)
\label{eq:cost}
\end{equation}
where \( l_i \) is the total number of tokens (prompt and response) for request \( r_i \), and \( p_{m_k} \) is the cost per million tokens\footnote{We refer to the token pricing from Together.ai for cost calculation purposes; their API was not used. \url{https://www.together.ai/pricing}} for the LLM \( m_k \) assigned to \( r_i \), as adopted in \cite{Feng2024, Stripelis2024}. This metric accounts for the computational and monetary cost of processing requests, which varies depending on the selected LLM (e.g., smaller edge-based models like qwen2.5:1.5b-instruct are cheaper than cloud-based models like gemma3:27b).

\subsection{\texorpdfstring{Response Time Metric (\( RT \))}{Response Time Metric (RT)}}
The response time metric \( RT \) captures the average end-to-end latency across all requests, defined as:
\begin{equation}
RT = \frac{1}{I} \sum_{i=1}^I RT_i
\label{eq:rt}
\end{equation}
where \( RT_i \) is the response time for request \( r_i \), calculated as:

\begin{align}
RT_i &= \left( \frac{Q_{\text{size},i}}{B_{\text{r}\to\text{j}}} + \text{latency}_{\text{r}\to\text{j}} \right) \notag \\
     &\quad + T_{\text{infer},i} + \left( \frac{R_{\text{size},i}}{B_{\text{j}\to\text{r}}} + \text{latency}_{\text{j}\to\text{r}} \right)
\label{eq:rt_i}
\end{align}

Here, \( Q_{\text{size},i} \) and \( R_{\text{size},i} \) are the sizes (in bytes) of the query and response for request \( r_i \), respectively. \( B_{\text{r}\to\text{j}} \) and \( B_{\text{j}\to\text{r}} \) denote the bandwidth (in bytes per second) from the router to node \( n_j \) and from node \( n_j \) to the router, respectively. \( \text{latency}_{\text{r}\to\text{j}} \) and \( \text{latency}_{\text{j}\to\text{r}} \) represent the network latency for data transfer to and from node \( n_j \). \( T_{\text{infer},i} \) is the inference time for request \( r_i \) on the assigned LLM, extracted from the LLM’s response metadata (e.g., JSON format).

This formulation captures the total latency, including query transmission, inference processing, and response transmission, accounting for both network conditions and computational efficiency of the assigned node-LLM pair.

\subsection{Discussion}

The multi-objective optimization problem defined in \eqref{eq:objective} is addressed using NSGA-II, which efficiently approximates the Pareto front of solutions to balance the competing objectives of response quality (\( RQ \)), inference cost (\( C \)), and response time (\( RT \)). NSGA-II employs non-dominated sorting to rank solutions based on their dominance, ensuring that solutions offering better trade-offs across all three objectives are prioritized \cite{ma2023comprehensive}. Additionally, the crowding distance mechanism promotes diversity within the Pareto front by favoring solutions in less populated regions of the objective space, preventing convergence to a narrow set of trade-offs \cite{squillero2016divergence}. This approach effectively handles the heterogeneity of requests (e.g., varying prompt lengths and complexity) and nodes (e.g., edge vs. cloud resources), enabling the routing policy to adapt to dynamic workloads and resource differences, thereby achieving a balanced optimization of the three evaluation metrics.

\section{Proposed Solution}

In this section, we present the design of our proposed routing algorithm, which leverages the Non-dominated Sorting Genetic Algorithm II (NSGA-II) to optimize the assignment of Large Language Model (LLM) inference requests across a cloud-edge architecture. The algorithm aims to minimize response quality metric (\( RQ \)), inference cost (\( C \)), and response time (\( RT \)), as defined in Equations \eqref{eq:rq}, \eqref{eq:cost}, and \eqref{eq:rt}, respectively, while adhering to the resource and assignment constraints outlined in Section III. We describe the system architecture, the NSGA-II-based routing mechanism, and the implementation details for dynamic workload management in a heterogeneous cloud-edge environment.

\subsection{System Architecture}

\begin{figure}[htbp]
    \centering
    \includegraphics[width=0.48\textwidth]{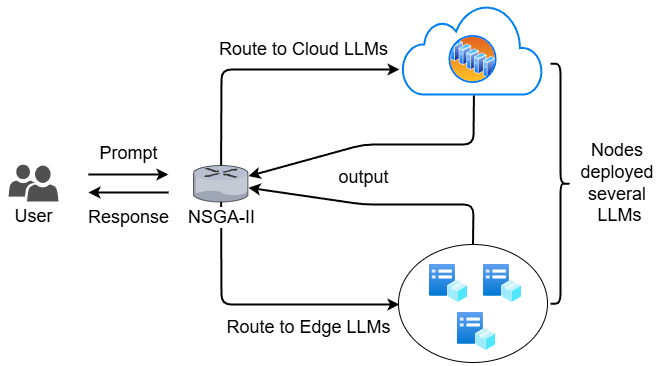}
    \caption{System architecture comprising cloud and edge nodes with request routing via NSGA-II.}
    \label{fig:system_architecture}
\end{figure}

Figure~\ref{fig:system_architecture} illustrates the system architecture, which operates within a cloud-edge framework comprising \( N \) nodes (\( N = N_{\text{cloud}} \cup N_{\text{edge}} \)). Each node \( n_j \in N \) hosts a subset of LLMs \( \mathcal{M}(n_j) \subseteq M \), as defined in Section~III. Edge nodes, typically resource-constrained devices, are deployed closer to users to reduce network latency, while cloud nodes, equipped with high-performance computing resources, offer greater capacity for complex requests. A router receives inference requests \( R = \{r_1, r_2, \dots, r_I\} \), each characterized by attributes such as prompt length, complexity, and category. The router employs the NSGA-II algorithm to assign each request \( r_i \) to a node-LLM pair \( (n_j, m_k) \in S \), where \( m_k \in \mathcal{M}(n_j) \), optimizing the multi-objective function in Equation \eqref{eq:objective}.

\subsection{NSGA-II-Based Routing Algorithm}
The NSGA-II algorithm is employed to find a Pareto-optimal set of routing policies \( \pi: R \to S \), balancing the trade-offs among \( RQ \), \( C \), and \( RT \). NSGA-II is well-suited for this task due to its ability to handle multi-objective optimization through non-dominated sorting and crowding distance mechanisms \cite{ma2023comprehensive}. The algorithm iteratively evolves a population of candidate routing policies to approximate the Pareto front, ensuring diverse solutions that address varying application priorities (e.g., low latency for real-time tasks or low cost for batch processing).

\subsubsection{Population Initialization}
The algorithm initializes a population of \( P \) candidate routing policies, where each policy \( \pi_p \) (for \( p = 1, 2, \dots, P \)) is a mapping \( \pi_p: R \to S \). Each policy assigns every request \( r_i \in R \) to a valid node-LLM pair \( (n_j, m_k) \in S \), where \( m_k \in \mathcal{M}(n_j) \). To ensure feasibility, assignments are constrained by the resource capacity \( C_j \) of each node \( n_j \), as defined in Section III. Initial policies are generated randomly, with a heuristic bias toward assigning lightweight requests (e.g., short prompts) to edge nodes and complex requests (e.g., long prompts or high-complexity tasks) to cloud nodes, based on model capabilities and node resources.

\subsubsection{Fitness Evaluation}
For each routing policy \( \pi_p \), the fitness is evaluated based on the three objectives:
\begin{itemize}
    \item \textbf{Response Quality (\( RQ \))}: Computed using Equation \eqref{eq:rq}, where the quality score \( q(r_i) \) is derived from dataset-specific metrics (e.g., F1-score for SQuAD, CodeBLEU for MBPP) for the assigned LLM \( m_k \).
    \item \textbf{Inference Cost (\( C \))}: Calculated using Equation \eqref{eq:cost}, based on the token count \( l_i \) of request \( r_i \) and the cost per million tokens \( p_{m_k} \) of the assigned LLM.
    \item \textbf{Response Time (\( RT \))}: Determined using Equation \eqref{eq:rt}, incorporating network transmission times, inference time \( T_{\text{infer},i} \), and node-specific latency and bandwidth parameters.
\end{itemize}
The fitness values are normalized to ensure comparability across objectives, using min-max normalization based on observed ranges from historical data or initial simulations.

\subsubsection{Non-Dominated Sorting and Crowding Distance}
NSGA-II employs non-dominated sorting to rank policies based on their dominance. A policy \( \pi_p \) dominates \( \pi_q \) if it is better in at least one objective and no worse in others. Policies are grouped into Pareto fronts, with the first front containing non-dominated solutions. To maintain diversity, NSGA-II computes the crowding distance for each policy within a front, measuring the Euclidean distance between neighboring solutions in the objective space (\( RQ \), \( C \), \( RT \)). Policies in less crowded regions are prioritized to ensure a well-distributed Pareto front \cite{squillero2016divergence}.

\subsubsection{Selection, Crossover, and Mutation}
Parent policies are selected using binary tournament selection, where two policies are compared, and the one with a higher rank (or higher crowding distance within the same rank) is chosen. Crossover is performed by combining assignments from two parent policies, swapping node-LLM pairs for a subset of requests to create offspring policies. Mutation introduces random changes by reassigning a small fraction of requests to different valid node-LLM pairs, ensuring exploration of the solution space while respecting resource constraints.

\subsubsection{Algorithm Workflow}
The NSGA-II workflow is summarized in Algorithm \ref{alg:nsga-ii}. The algorithm iterates for \( T \) generations, updating the population by combining parent and offspring policies and selecting the top \( P \) solutions based on non-dominated sorting and crowding distance. Upon convergence, the algorithm outputs a Pareto-optimal set of routing policies, from which a specific policy can be selected based on application-specific weights \( \omega_1, \omega_2, \omega_3 \) (Equation \eqref{eq:objective}) or user priorities.

\begin{algorithm}
\caption{NSGA-II workflow}
\label{alg:nsga-ii}
\begin{algorithmic}[1]
    \STATE Initialize population \( P \) of \( P \) routing policies \( \pi_p: R \to S \)
    \FOR{each generation \( t = 1 \) to \( T \)}
        \STATE Evaluate fitness of \( P \) using \( RQ \), \( C \), and \( RT \) (Equations \eqref{eq:rq}, \eqref{eq:cost}, \eqref{eq:rt})
        \STATE Perform non-dominated sorting to rank policies
        \STATE Compute crowding distance for diversity
        \STATE Select parents via binary tournament selection
        \STATE Apply crossover to generate offspring \( Q \)
        \STATE Apply mutation to \( Q \), ensuring resource constraints
        \STATE Combine \( P \) and \( Q \), select top \( P \) policies for next generation
    \ENDFOR
    \STATE Output Pareto-optimal set of routing policies
\end{algorithmic}
\end{algorithm}

\subsubsection{Routing Policy Execution}
The routing policy \( \pi^* \), selected from the Pareto-optimal set generated by NSGA-II, maps each inference request \( r_i \in R \) to a node-LLM pair \( (n_j, m_k) \in S \). The execution of \( \pi^* \) involves three key steps: feature extraction, rule-based assignment, and dynamic policy updates.

\textbf{Feature Extraction.} For each request \( r_i \), the router extracts a feature vector \( \mathbf{f}_i = (c_i, t_i, q_j) \), where:
\begin{itemize}
    \item \( c_i \): Complexity score, computed as a weighted combination of features extracted from the prompt, including token length, sentence count, task type, and presence of output constraints (e.g., `must', `only', `output'). Weights are empirically tuned based on correlations between features and inference time from training data and the score is normalized to \([0, 1]\).
    \item \( t_i \): Task category (e.g., `code', `math', `general') and confidence score, predicted by a SetFit-based classifier\footnote{\url{https://github.com/huggingface/setfit}} trained on samples from datasets (SQuAD, GSM8K, HellaSwag and MBPP).
    \item \( q_j \): Current queue length of node \( n_j \), obtained from the monitoring module.
\end{itemize}

\textbf{Rule-Based Routing.} The router applies a set of rules derived from NSGA-II-optimized thresholds to assign \( r_i \) to a node-LLM pair \( (n_j, m_k) \), as summarized in Algorithm~\ref{alg:runtime-routing}:
\begin{itemize}
    \item \textbf{Node Type Selection}: The complexity score \( c_i \), computed based on the prompt and task type, is compared against a task-specific difficulty threshold: if \( c_i < \theta_{d,\text{code}} \) for task type `code', \( c_i < \theta_{d,\text{math}} \) for `math', or \( c_i < \theta_{d,\text{general}} \) for other tasks (e.g., \( \theta_{d,\text{general}} = 0.5 \)), route to an edge node; otherwise, route to the model on the cloud node.
    \item \textbf{Edge Node Selection}: From edge nodes, select candidates where the current queue length \( q_j \leq \theta_q \) (queue threshold, e.g., 5 requests); if no candidates are available, fall back to a high-capacity model like gemma3:27b on a cloud node.
    \item \textbf{Model Selection}: If the classifier predicts task category \( t_i \) (e.g., `code') with confidence \( p_t \geq \theta_{t,\text{code}} \) (e.g., 0.7) or \( t_i = \) `math' with \( p_t \geq \theta_{t,\text{math}} \), select a specialized model type (e.g., `coder' for code, `math' for math); otherwise, select a general model type (e.g., `instruct'). The first available edge node hosting a model matching the selected type is chosen.
\end{itemize}
These thresholds (\( \theta_{d,\text{code}}, \theta_{d,\text{math}}, \theta_{d,\text{general}}, \theta_q, \theta_{t,\text{code}}, \theta_{t,\text{math}} \)) are optimized offline by NSGA-II, treating them as decision variables in the population, evaluated against response quality (\( RQ \)), inference cost (\( C \)), and response time (\( RT \)), as defined in Equations \eqref{eq:rq}, \eqref{eq:cost}, and \eqref{eq:rt}. During actual request routing, the policy \( \pi^* \) is implemented as a static lookup table that maps request features \( \mathbf{f}_i \) to node-LLM pairs, enabling millisecond-level routing decisions. To adapt to dynamic workloads, small-scale NSGA-II re-optimization can be triggered periodically (e.g., every hour or once per day).

\begin{algorithm}
\caption{Runtime LLM Request Routing}
\label{alg:runtime-routing}
\begin{algorithmic}[1]
    \STATE \textbf{Input}: Request \( r_i \), policy \( \pi^* \), system state, thresholds
    \STATE Extract features \( \mathbf{f}_i = (c_i, t_i, q_j) \)
    \STATE Predict task type \( t_i \) and confidence \( p_t \) via classifier
    \STATE Compute complexity score \( c_i \) based on prompt and \( t_i \)
    \IF{\( t_i = \) `code' and \( c_i < \theta_{d,\text{code}} \)}
        \STATE go\_edge = true
    \ELSIF{\( t_i = \) `math' and \( c_i < \theta_{d,\text{math}} \)}
        \STATE go\_edge = true
    \ELSIF{\( c_i < \theta_{d,\text{general}} \)}
        \STATE go\_edge = true
    \ELSE
        \STATE go\_edge = false
    \ENDIF
    \IF{go\_edge}
        \STATE Filter edge nodes where \( q_j \leq \theta_q \)
        \IF{no candidates available}
            \STATE Select the high-capacity model on cloud node
        \ELSE
            \IF{\( t_i = \) `code' and \( p_t \geq \theta_{t,\text{code}} \)}
                \STATE Select model type `coder'
            \ELSIF{\( t_i = \) `math' and \( p_t \geq \theta_{t,\text{math}} \)}
                \STATE Select model type `math'
            \ELSE
                \STATE Select model type `instruct'
            \ENDIF
            \STATE Select first edge node with matching model
        \ENDIF
    \ELSE
        \STATE Select the high-capacity model on cloud node
    \ENDIF
    \STATE Forward \( r_i \) to node \( n_j \), model \( m_k \)
    \STATE \textbf{Output}: Assigned pair \( (n_j, m_k) \)
\end{algorithmic}
\end{algorithm}


\subsection{Discussion}
Our NSGA-II-based routing algorithm effectively addresses the challenges of request heterogeneity and node diversity in cloud-edge systems. By optimizing across \( RQ \), \( C \), and \( RT \), it ensures high-quality responses for diverse tasks (e.g., SQuAD, MBPP) while minimizing latency and cost. The periodic re-optimization enables adaptability to dynamic conditions, making the algorithm suitable for real-world LLM deployments. In the next section, we describe the experimental setup to validate the algorithm’s performance using benchmark datasets and realistic cloud-edge scenarios.


\section{Experiment}

This section evaluates our NSGA-II-based routing algorithm in a real cloud-edge environment, focusing on its performance in terms of response quality (\( RQ \)), response time (\( RT \)), and inference cost (\( C \)). We describe the implementation, experimental setup, datasets, and comparative analysis with baseline methods.

\subsection{Implementation}
The routing algorithm is implemented in Python, utilizing tiktoken\footnote{\url{https://github.com/openai/tiktoken}} for token counting and Ollama\footnote{\url{https://ollama.com}} for LLM inference. The NSGA-II optimization is performed using the pymoo library\footnote{\url{https://github.com/anyoptimization/pymoo}}, with parameters set as follows: population size \( P = 100 \), number of generations \( T = 100 \), crossover probability 0.8, and mutation probability 0.1. These parameters are tuned to balance exploration and exploitation while maintaining computational efficiency.

\subsection{Datasets}

To evaluate LLM performance across diverse tasks in real-world settings, we employ four widely used datasets, mixed rather than tested sequentially. The \textit{MBPP} dataset~\cite{austin2021program} contains 974 introductory programming problems with text descriptions and test cases, focusing on code generation. \textit{GSM8K}~\cite{zhang2024careful} includes 8,498 grade-school-level math word problems with detailed solution steps, testing multi-step arithmetic and algebraic reasoning. \textit{SQuAD}~\cite{rajpurkar2016squad} consists of approximately 100,000 question-answer pairs from Wikipedia, targeting reading comprehension and precise answer extraction. Lastly, \textit{HellaSwag}~\cite{zellers2019hellaswag} presents 70,000 everyday scenario contexts with one correct and three adversarial endings, assessing commonsense reasoning and contextual understanding.

\subsection{Experimental Setup}
The system is deployed on a heterogeneous cloud-edge environment with three resource-constrained edge nodes (4-core CPU, 8GB RAM, no GPU) and one high-performance cloud node (60-core CPU, NVIDIA A40 24GB GPU). The cloud node hosts the model `gemma3:27b', while each edge node deploys `qwen2.5:1.5b-instruct', `qwen2.5-coder:1.5b-instruct', and `qwen2.5-math:1.5b-instruct', reflecting model heterogeneity.

A test script mixes problems from MBPP, GSM8K, SQuAD, and HellaSwag, sending 500 requests in total with a round-robin order (e.g., MBPP, GSM8K, HellaSwag, SQuAD, repeating). The requests are evenly distributed across the four datasets, with 125 requests per dataset. Each request is logged with attributes including dataset, global index, assigned model, node IP, prediction, inference quality, reference answer, response time (in seconds), and cost\footnote{We refer to the token pricing from Together.ai for cost calculation purposes; their API was not used. \url{https://www.together.ai/pricing}} (in US dollars).

To our knowledge, there is currently no work that simultaneously considers the routing problem of multiple nodes in a cloud-edge environment where heterogeneous models are deployed on the nodes. Therefore, we consider the following baselines:
\begin{itemize}
    \item \textbf{Cloud Only}: All requests are routed to the cloud node with \texttt{gemma3:27b}.
    \item \textbf{Edge Only}: All requests are routed to one of the models deployed on an edge node according to the request type.
    \item \textbf{Random Router}: Requests are randomly routed to any node-LLM pair.
    \item \textbf{Round Robin Router}: Requests are evenly routed to cloud and edge nodes in a cyclic manner, model selection based on the request type.
\end{itemize}

\subsection{Results and Analysis}

\begin{figure}[htbp]
    \centering
    \includegraphics[width=0.5\textwidth]{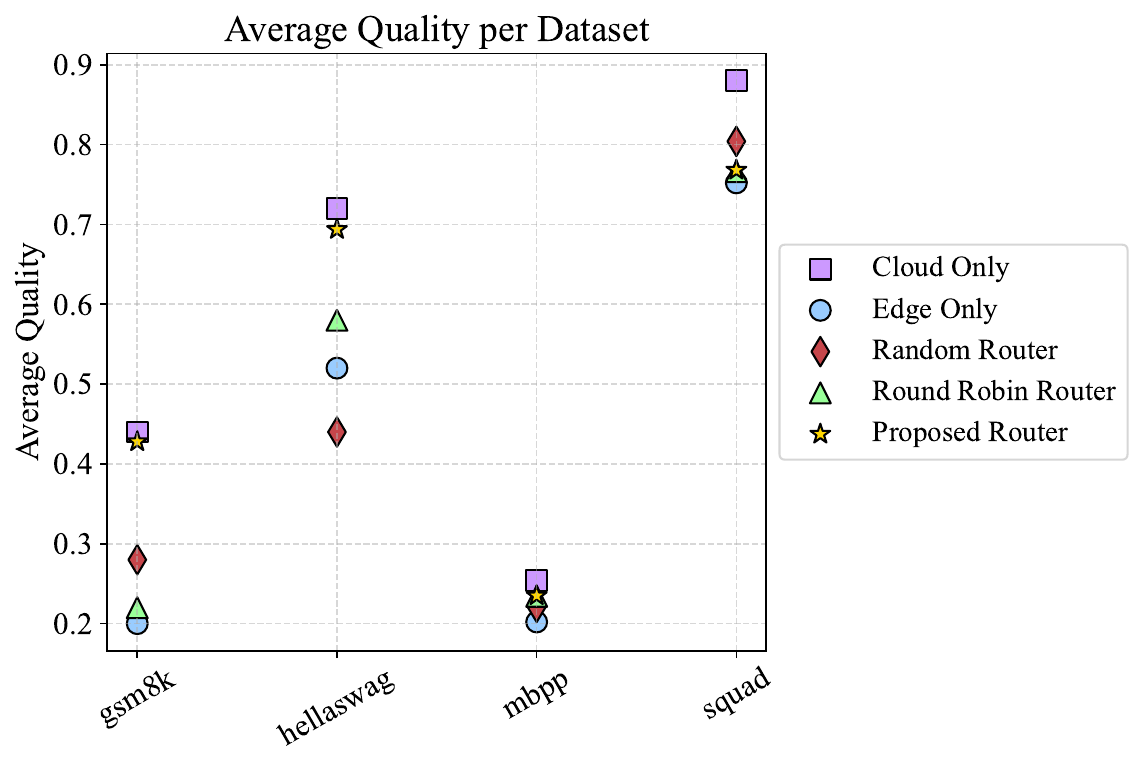}
    \caption{Average quality per dataset for different routing strategies. Our solution achieves competitive quality across MBPP, GSM8K, SQuAD, and HellaSwag.}
    \label{fig:quality_per_dataset}
\end{figure}

\begin{figure}[htbp]
    \centering
    \includegraphics[width=0.46\textwidth]{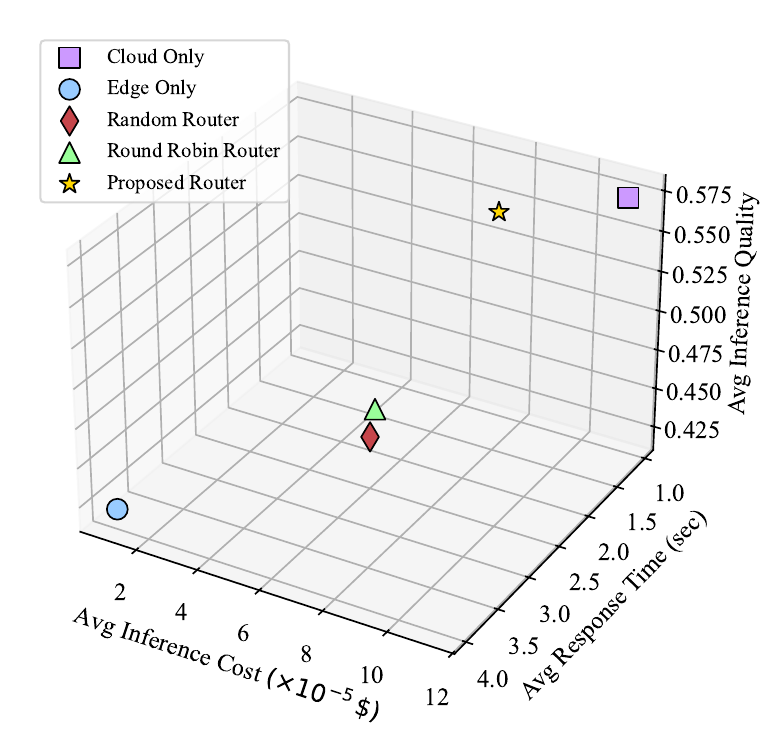}
    \caption{Three-dimensional performance comparison of routing strategies, illustrating trade-offs among model performance (quality), response time, and cost.}
    \label{fig:performance_3d}
\end{figure}

\begin{figure*}[htbp]
    \centering
    \includegraphics[width=0.9\textwidth]{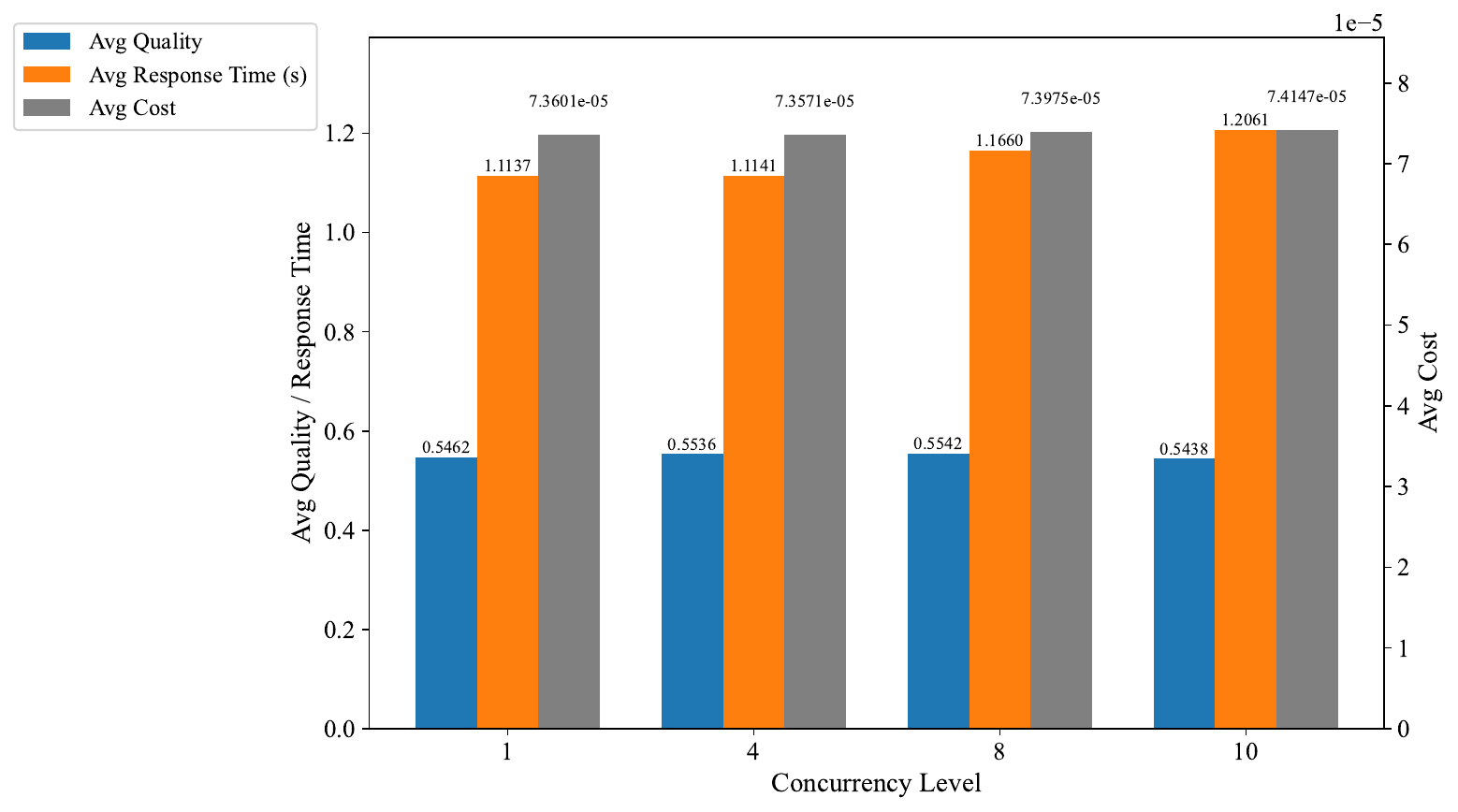}
    \caption{Performance of our proposed router at different concurrency levels (1, 4, 8, 10) in terms of average quality, response time, and cost.}
    \label{fig:concurrency_analysis}
\end{figure*}

The performance of our NSGA-II-based router and baseline methods is evaluated based on response quality (\( RQ \)), response time (\( RT \)), and inference cost (\( C \)). Results are summarized in Table \ref{tab:comparison} and visualized in Figures \ref{fig:quality_per_dataset}, \ref{fig:performance_3d}, and \ref{fig:concurrency_analysis}.

\begin{table*}[htbp]
\centering
\caption{Performance comparison of different routing strategies}
\label{tab:comparison}
\renewcommand{\arraystretch}{1.3}
\setlength{\tabcolsep}{4.5pt}
\begin{tabular}{lcccc}
\toprule
\textbf{Router} & \textbf{avg\_quality ($\uparrow$)} & \textbf{avg\_response\_time ($\downarrow$)} & \textbf{avg\_cost ($\downarrow$)} & \textbf{overall ($\uparrow$)} \\
\midrule
Cloud Only           & 0.5736 & 1.0624 & 1.13e-4 & 0.6667 \\
Edge Only            & 0.4207 & 3.9673 & 9.00e-6 & 0.3333 \\
Random Router        & 0.4361 & 2.3571 & 5.71e-5 & 0.3834 \\
Round Robin Router   & 0.4618 & 2.4971 & 6.16e-5 & 0.4062 \\
\textbf{Proposed Router} & 0.5462 & 1.1137 & 7.36e-5 & \textbf{0.7523} \\
\bottomrule
\end{tabular}
\end{table*}

Table \ref{tab:comparison} summarizes five routing strategies across four metrics:
\begin{itemize}
  \item \textbf{avg\_quality} ($\uparrow$): the average response quality on a 0--1 scale (higher is better);
  \item \textbf{avg\_response\_time} ($\downarrow$): the average response time in seconds (lower is better);
  \item \textbf{avg\_cost} ($\downarrow$): the average inference cost per request in US dollars (lower is better);
  \item \textbf{overall} ($\uparrow$): a composite score in $[0,1]$ that equally weights normalized quality, latency, and cost (higher is better).
\end{itemize}

\textbf{Overall} is for quantitatively assessing the trade-offs among the three metrics. Specifically, we first define the following:
\begin{align*}
Q_i &= \text{avg\_quality of router } i, \\
T_i &= \text{avg\_response\_time of router } i, \\
C_i &= \text{avg\_cost of router } i\,.
\end{align*}
for $i=1,\dots,5$ (the five strategies).  Then let
\[
  Q_{\min} = \min_i Q_i,\quad Q_{\max} = \max_i Q_i,
\]
\[
  T_{\min} = \min_i T_i,\quad T_{\max} = \max_i T_i,
\]
\[
  C_{\min} = \min_i C_i,\quad C_{\max} = \max_i C_i.
\]
Next, normalize each dimension so that “larger = better”:
\[
\begin{gathered}
  q_{i}^{\mathrm{norm}} 
  = \frac{\,Q_i - Q_{\min}\,}{\,Q_{\max} - Q_{\min}\,}, \\[6pt]
  \begin{aligned}
    t_{i}^{\mathrm{norm}} 
    &= \frac{\,T_{\max} - T_i\,}{\,T_{\max} - T_{\min}\,},
    &\quad
    c_{i}^{\mathrm{norm}} 
    &= \frac{\,C_{\max} - C_i\,}{\,C_{\max} - C_{\min}\,}
  \end{aligned}
\end{gathered}
\]

Finally, we compute the composite score as the mean of the three normalized metrics:
\[
  \text{overall}_i 
  = \frac{1}{3}\bigl(q_{i}^{\mathrm{norm}} + t_{i}^{\mathrm{norm}} + c_{i}^{\mathrm{norm}}\bigr).
\]

Based on the metrics and the resulting overall scores, we now examine how each routing strategy performs. From the data in Table \ref{tab:comparison}, the “Cloud Only” strategy achieves the highest average quality (0.5736) and lowest latency (1.0624\,s) but also the highest cost (1.13\,$\times\,10^{-4}$\,\$), yielding an overall of 0.6667. In contrast, “Edge Only” minimizes cost (9.00\,$\times\,10^{-6}$\,\$) at the expense of quality (0.4207) and response time (3.9673\,s), resulting in an overall of 0.3333. “Random Router” and “Round Robin Router” exhibit middling performance in all three dimensions, with overall scores of 0.3834 and 0.4062, respectively. Our “Proposed Router” attains average quality = 0.5462, average response\_time = 1.1137\,s, average cost = 7.36\,$\times\,10^{-5}$\,\$, which normalize to approximately $q_{\text{norm}}=0.8703$, $t_{\text{norm}}=0.9780$, $c_{\text{norm}}=0.4087$, yielding the highest overall of 0.7523. This confirms that our proposed router achieves the best aggregate trade‐off among quality, latency, and cost.

Figure \ref{fig:quality_per_dataset} highlights dataset-specific performance, where our solution excels on SQuAD (approaching 0.9) and MBPP (around 0.7), benefiting from task-specific model assignments (e.g., `qwen2.5-coder:1.5b-instruct' for MBPP). Edge Only and Random Router exhibit lower quality across all datasets (e.g., below 0.5 for GSM8K), reflecting their inability to adapt to task complexity. Cloud Only maintains high quality (e.g., 0.9 for SQuAD), but our proposed router's performance is competitive, suggesting effective resource allocation.

Figure~\ref{fig:performance_3d} presents a three-dimensional visualization of the trade‐offs among average inference quality (Z‐axis), response time (Y‐axis), and cost (X‐axis). Ideally, a routing strategy should lie near the \emph{upper‐right edge} of the quality–latency plane while remaining as far left as possible on the cost axis.

By comparing these points in three dimensions, it is clear that our proposed router achieves the best overall trade‐off: it is nearly as far upward (high quality) and backward (low latency) as Cloud Only, while remaining much farther left (lower cost). Key characteristics of each strategy are summarized as follows:

\begin{itemize}
  \item \textbf{Proposed Router}: Inference quality $\approx 0.54$, response time $\approx 1.1\ \mathrm{s}$, cost $\approx 7\times10^{-5}\ \$$. It lies almost as high and as far back as Cloud Only (indicating comparable quality and latency), yet much farther left (indicating significantly lower cost), thereby achieving the optimal overall trade‐off.
  \item \textbf{Cloud Only}: Inference quality $\approx 0.58$ (highest) but cost $\approx 1.13\times10^{-4}\ \$$ (highest). It appears at the “highest” and “farthest back” position, but is shifted rightward, indicating a higher cost.
  \item \textbf{Edge Only}: Cost $\approx 9\times10^{-6}\ \$$ (lowest), corresponding to the “farthest left” point; however, its inference quality $\approx 0.42$ (lowest) and response time $\approx 3.97\ \mathrm{s}$ (highest) place it at the “lowest” and “nearest” region, resulting in a weaker overall performance.
  \item \textbf{Random Router} and \textbf{Round Robin Router}: These baselines occupy intermediate positions. They cannot match Cloud Only’s high quality and low latency, nor Edge Only’s ultra‐low cost, and thus do not outperform our Proposed Router in the overall trade‐off.
\end{itemize}

In addition, we evaluate our proposed router's performance under varying concurrency levels to assess scalability. The experiment begins with a concurrency equal to the number of nodes (4), then increasing to 8 and 10. Figure~\ref{fig:concurrency_analysis} illustrates the performance of our proposed router under concurrency levels 1, 4, 8, and 10. At a concurrency of 1, the router achieves baseline performance with a quality of 0.5462, response time of 1.1137\,s, and a cost of 7.36\,$\times\,10^{-5}$\,\$. Increasing the concurrency to 4 slightly improves quality to 0.5536, with marginal increments in response time (1.1141\,s) and negligible changes in cost (7.36\,$\times\,10^{-5}$\,\$), indicating effective load distribution at moderate concurrency. As concurrency rises further to 8, quality stabilizes at 0.5542, though response time (1.1660\,s) and cost (7.40\,$\times\,10^{-5}$\,\$) begin to show moderate increases, suggesting emerging resource contention. At a concurrency level of 10, system capacity constraints become more pronounced, resulting in a reduced quality of 0.5438, elevated response time of 1.2061\,s, and higher cost of 7.41\,$\times\,10^{-5}$\,\$. These results highlight the robustness and scalability of our proposed router under moderate concurrency conditions.

\subsection{Discussion}
The NSGA-II-based routing algorithm demonstrates robust optimization of response quality (\( RQ \)), response time (\( RT \)), and inference cost (\( C \)) within a heterogeneous cloud-edge environment. Its superior performance over baseline methods stems from adaptive assignment strategies that account for request complexity and node availability, as validated by the mixed-dataset workload comprising 500 requests. The algorithm's scalability is further evidenced by its stable performance across varying concurrency levels (up to 10), with degradation only observed near the system's capacity limit (concurrency 11).

\section{Conclusion}

This paper addresses the critical challenge of efficiently routing inference requests for Large Language Models (LLMs) within a cloud-edge environment, responding to the growing demand for scalable and cost-effective inference services. We formulated the inference routing problem as a multi-objective optimization task, balancing response quality (\(RQ\)), response time (\(RT\)), and inference cost (\(C\)). To address this, we introduced a novel routing algorithm leveraging the Non-dominated Sorting Genetic Algorithm II (NSGA-II), enabling adaptive assignment of heterogeneous inference requests---varying in complexity, prompt length, and category---to diverse LLM instances across edge and cloud nodes. Additionally, to support the evaluation of our method, we also developed a set of performance testing tools tailored for distributed cloud-edge environments.

Experimental evaluation on a realistic cloud-edge testbed using benchmarks such as SQuAD, MBPP, HellaSwag, and GSM8K demonstrated our method's robustness and effectiveness. Specifically, our router achieved an average response quality of 0.5462, attaining 95.2\% of the Cloud-Only baseline (0.5736), with only a modest 4.8\% increase in response time (1.1137 vs. 1.0624 seconds). Significantly, inference costs were reduced by approximately 34.9\% (from 1.13\,$\times\,10^{-4}$\,\$ to 7.36\,$\times\,10^{-5}$\,\$). Compared with baseline approaches such as Edge Only, Random Router, and Round Robin Router, our method excelled in efficiently balancing trade-offs among response quality, latency, and cost, maintaining consistent performance across varying concurrency scenarios.

This research provides a foundational framework for adaptive LLM inference routing in heterogeneous cloud-edge environments, significantly advancing the practical deployment of LLM services, especially in resource-constrained and dynamic contexts.

\textbf{Future Work:} To further enhance our approach, we plan to integrate comprehensive real-time monitoring mechanisms for node and model status, coupled with fault-tolerant strategies to improve reliability under node or model failures. Additionally, as inference requests increasingly shift towards multimodal formats---including text, image, and audio---we aim to extend our method to support multimodal inference routing, broadening the applicability and robustness of the proposed solution.

\section{Ethics and Data Privacy}

Our work relies exclusively on publicly available benchmark datasets: SQuAD, MBPP, HellaSwag, and GSM8K, as well as programmatically generated inference requests and theoretical simulations. No human participants or real‐world user data were involved, and no personally identifiable or sensitive information was collected, stored, or processed. All datasets used are fully anonymized and contain no names, contact details, or other private attributes. Consequently, no ethical approval or data privacy review was required.





\bibliographystyle{IEEEtran}
\bibliography{references}

\end{document}